\documentclass[twocolumn]{webofc}
\usepackage[varg]{txfonts}   
\begin{document}
\title{Superfluid effects in collision between systems with small particle number
}

\author{Guillaume Scamps\inst{1}\fnsep\thanks{\email{scamps@nucl.ph.tsukuba.ac.jp}} \and
        Yukio Hashimoto \inst{1}\fnsep\thanks{\email{hashimoto@nucl.ph.tsukuba.ac.jp}} 
}

\institute{ 
           Center for Computational Sciences, University of Tsukuba, Tsukuba 305-8571, Japan
          }

\abstract{  The interpretation of the new effect of the superfluidity in reactions with small number of particles is discussed in a simple model where the exact solution is accessible. It is find that the fluctuations of observable with the gauge angle reproduce well the exact fluctuations. Then a method of projection is proposed and tested to determine the transfer probabilities between two  superfluid systems.}
\maketitle
\section{Introduction}
\label{intro}

In a precedent contribution \cite{Has16}, we studied the reaction between two superfluid nuclei  $^{20}$O+$^{20}$O with the Time-dependent Hartree-Fock-Bogoliubov theory. The dependence of the observables with respect to the initial relative gauge angle between the two initial fragments have been studied. A Josephson effect have been found, as well as a dependence of the nucleus-nucleus potential with this gauge angle. Those effects have also been studied in Ref. \cite{Mag16}. As point out by Ref. \cite{Bul17}, the interpretation of this result is ambiguous for systems which should have a given number of particles in each fragments initially and so respect the gauge angle symmetry. This motivate us to consider a more simple model that can be solved exactly.

\section{Model}

Our model is inspired by the Dietrich model of nuclear Josephson effect \cite{Die70}. Two systems are initially considered ${\cal S}_1$ and ${\cal S}_2$ both are composed of $\Omega$ doubly degenerated level with $N$ particles. The initial hamiltonian of the system is composed of a pairing interaction,
\begin{align}
\hat H_{0} = G \sum_{ij \in {\cal S}_1} \hat a^{\dagger}_{ i} \hat a^{\dagger}_{\bar i}  \hat a_{ \bar j} \hat a_{ j}  + G \sum_{ij \in {\cal S}_2} \hat a^{\dagger}_{ i} \hat a^{\dagger}_{\bar i}  \hat a_{ \bar j} \hat a_{ j}.
\end{align}
For simplicity, we consider all the particles paired. The exact solution of this hamiltonian is obtain by diagonalizing the hamiltonian in the space of all the configuration that have $N$ particles in both sides. We consider the case of $\Omega$=4, $N$=4 and $G=-0.2$ MeV. In that case, the total energy of the system with the exact solution is 1.6 MeV. For the HFB solution, we adjust the interaction $G^{\rm eff}$=$\frac43 G$ in order to obtain the same initial energy.
 
The initial state is then propagated in time, with the time-dependent hamiltonian that connect the two systems,
\begin{align}
\hat H(t) = \hat H_{0} + V(t) \sum_{i \in {\cal S}_1 ; j \in {\cal S}_2 }  \left(   \hat a^{\dagger}_{ i} \hat a^{\dagger}_{\bar i}  \hat a_{ \bar  j} \hat a_{ j}  +   \hat a^{\dagger}_{ j} \hat a^{\dagger}_{\bar j}  \hat a_{ \bar i} \hat a_{ i}  \right).
\end{align}
For simplicity reason, the time dependent interaction is chosen as,
\begin{align}
V(t) = V_0 e^{-at^2},
\end{align}
with $a$=0.3$\times 10^{44}$s$^{-2}$.
The system is evolved between time t=-15$\times 10^{-22}$s and  t=15$\times 10^{-22}$s. An effective interaction $V_0^{\rm eff}$=$\frac43 V_0$  is also taken in order to be consistent with the initial calculation. Two values of $V_0$ are tested, a weak and a strong interaction are chosen respectively with  $V_0=-0.003$ and $-0.4$ MeV. The two calculations, exact and TDHFB differs at the initialization of the calculation. For the exact case, there is no gauge angle, because the exact state has a good number of particles in both fragments so only one calculation is done. But for the TDHFB dynamic, the initial total state is defined as :
\begin{align}
|\Psi_{\varphi} \rangle =  \prod_{i \in {\cal S}_1} (u_i + v_i a^{\dagger}_{ i} a^{\dagger}_{\bar i} ) \prod_{j \in {\cal S}_2} ( u_j + e^{2i\varphi} v_j a^{\dagger}_{ j} a^{\dagger}_{\bar j} )   | ->,
\end{align} 
with $\varphi$ the relative gauge angle. So 24 initial values of $\varphi$ are chosen in the interval $[0,\pi]$.

\section{Results}

\begin{figure}[!ht]
\centering
\includegraphics[width= 0.9 \linewidth]{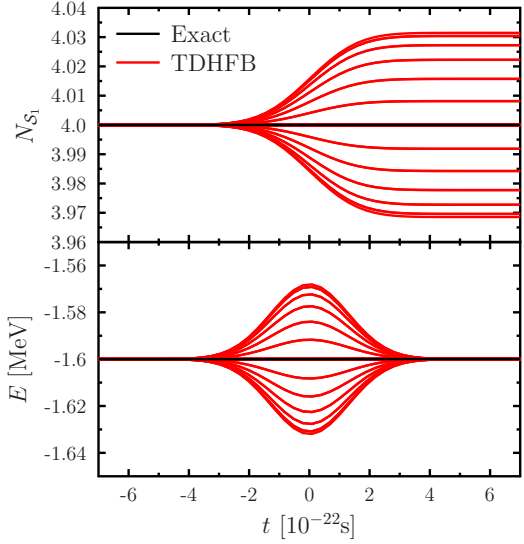}
\caption{ Comparison of the exact and TDHFB average number of particles in the ${\cal S}_1$ system (up) and the average energy (down) as a function of time in the weak interaction case. Each red lines correspond to a calculation with a given value of the initial relative gauge angle $\varphi$.  }
\label{fig:fluctEN_fct_t1_2}      
\end{figure}

Two observables are computed as a function of time, the number of particles in the ${\cal S}_1$ system $\langle \hat N_{{\cal S}_1} \rangle $ and the total energy $\langle \hat H(t) \rangle$. In this model, we expect to reproduce with TDHFB the Josephson effect and the fluctuations of the internal energy described in the introduction. Note that in this model, the energy is not conserved because the hamiltonian changes as a function of time. This is what we find in the fig. \ref{fig:fluctEN_fct_t1_2}. The TDHFB  average number of particles in one of the fragments changes with respect to $\varphi$ following a $\sin(2\varphi)$ dependence and the energy changes linearly  with  $\cos^2(\varphi)$. 

If we compare the average value of $\langle \Psi_{\varphi} | \hat {\cal O} | \Psi_{\varphi} \rangle $ we find a good agreement with the exact case. But it is more difficult to interpret the fluctuations of those observables. In that aim, we compute the quantal fluctuations of the observable  in the exact case,
\begin{align} 
	\sigma^{\rm Exact}_{\hat {\cal O}} = \sqrt{ \langle \Psi | \hat {\cal O}^2 | \Psi \rangle - \langle \Psi | \hat {\cal O} | \Psi \rangle^2 },
\end{align} 
to the standard deviation of the TDHFB observable with respect  to $\varphi$,
%
\begin{align} 
	\sigma^{\rm TDHFB}_{\hat {\cal O}} = \sqrt{  \frac1{\pi} \int_0^{\pi} \langle \Psi_{\varphi} | \hat {\cal O} | \Psi_{\varphi} \rangle^2  d\varphi  - \left( \frac1{\pi} \int_0^{\pi}  \langle \Psi_{\varphi} | \hat {\cal O} | \Psi_{\varphi} \rangle d\varphi \right)^2  }.
\end{align} 
%

\begin{figure}[!ht]
\centering
\includegraphics[width= 0.9 \linewidth]{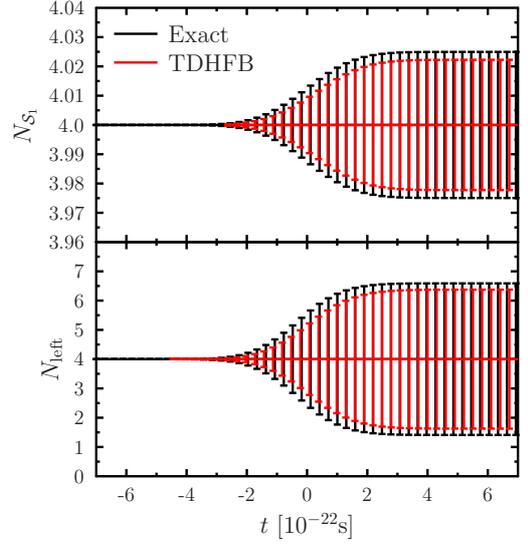}
\caption{ Comparison of the exact and TDHFB results for the number of particles in the ${\cal S}_1$ system as a function of time in the weak (up) and strong (down) interaction case. The error bars represent the fluctuations of this number. }
\label{fig:fluctN_fct_t3_28}      
\end{figure}

\begin{figure}[!ht]
\centering
\includegraphics[width= 0.9 \linewidth]{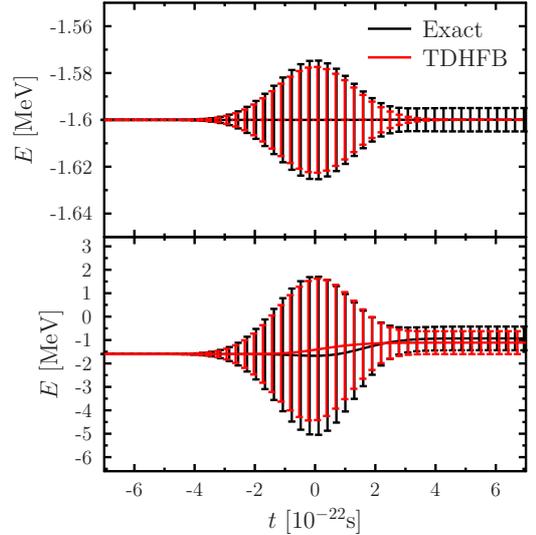}
\caption{ Comparison of the exact and TDHFB results for the energy as a function of time in the weak (up) and strong (down) interaction cases. The error bars represent the fluctuations of the energy. }
\label{fig:fluct_fct_t3_28}      
\end{figure}

These fluctuations are included in fig. \ref{fig:fluctN_fct_t3_28} and \ref{fig:fluct_fct_t3_28} by error bars. A very good agreement is found, showing that this interpretation of the fluctuations of the observable is a good one. Note that we should not take into account the direct calculation of the fluctuations of the observable in each trajectories,
\begin{align} 
	\sigma^{\varphi}_{\hat {\cal O}} = \sqrt{ \langle \Psi_{\varphi} | \hat {\cal O}^2 | \Psi_{\varphi} \rangle - \langle \Psi_{\varphi} | \hat {\cal O} | \Psi_{\varphi} \rangle^2 }.\label{eq:fluct_HFB}
\end{align} 
Indeed, at the initial time, these fluctuations are non zero because of the number of particles is not a good quantum number and because the HFB ground state is not an eigenstate of the hamiltonian. Then, to convolute the distribution of $\sigma^{\varphi}_{\hat {\cal O}}$ with  $ \sigma^{\rm TDHFB}_{\hat {\cal O}} $ will only bring spurious results. Then in realistic cases, the present interpretation should be used with caution and a more rigorous method of restauration of the gauge angle symmetry should be consider.

\section{Projection method}

Although fluctuations of observables brings interesting information, we would like to develop a method of projection in order to determine the complete distribution of particles in each fragments. To determine the transfer probabilities, we use the projection method \cite{Sim10} that determine the probability $P_{\cal S}(N)$ to have a given number of particles $N$ in a subspace $\cal S$,
\begin{align}
	P_{\cal S}(N,t) = \langle \Psi(t) | \hat P_{\cal S}(N) | \Psi(t) \rangle,
\end{align}
with 
\begin{align}
	\hat P_{\cal S}(N) = \frac{1}{2 \pi} \int_0^{2 \pi}  e^{i \varphi \hat N_{\cal S} - N} d\varphi.
\end{align}
$\hat N_{\cal S}$ is the operator that count the number of particles in the subspace $\cal S$. 

As discussed in Ref. \cite{Sca13}, the projection method has to be modified when used with theories that mix different number of particles. It is the case here, where we use quasi-particle states  that contain initially components  with the good number of particles $N_{\rm tot}$ but also components with ..., $N_{\rm tot}$-4, $N_{\rm tot}$-2,   $N_{\rm tot}$+2, $N_{\rm tot}$+4,... Those components will induce spurious results. For exemple, at the initial time, even if the two fragments are separated by an infinite distance, the pair transfer probability will not be zero.  

To suppress the spurious component, a double projection technique is applied,
\begin{align}
	P_{\cal S}(N) = \frac{ \langle \Psi(t) | \hat P_{\cal S}(N) \hat P(N_{\rm tot})  | \Psi(t) \rangle }{   \langle \Psi(t) |  \hat P(N_{\rm tot})  | \Psi(t) \rangle  } , \label{eq:dbl_prj}
\end{align}
with $\hat P(N_{\rm tot})$ the projector onto the total space.
This method only works for collisions where only one of the fragments is in the superfluid phase, the other one has to be in the normal phase.
Let's see what happen, if we use this method to collisions where both fragments break the gauge angle symmetry with average number $N_1$ and $N_2$ respectively for system 1 and 2. The total wave function after projection onto the good total number $N_{\rm tot}=N_1+N_2$ will still contain spurious components with an initial number of particles $N_1-2n$ and  $N_2+2n$ with $n$ an integer.
Those components will correspond to pair transfer states, and are spurious because they can happen before the collision.

Then for the collisions between two superfluid systems, we have to compute the probability
\begin{align}
	P_{\cal S}(N,t) = \frac{ \langle \Psi_{N_1,N_2}(t) | \hat P_{\cal S}(N) | \Psi_{N_1,N_2}(t) \rangle }{\langle \Psi_{N_1,N_2}(t) | \Psi_{N_1,N_2}(t) \rangle },
\end{align}	
with the state,
\begin{align}
| \Psi_{N_1,N_2}(t) \rangle  = \hat U^{TDPHFB}(t_0,t) \hat P_{\cal S}(N_1) \hat P_{\cal \bar S}(N_2) | \Psi(t_0) \rangle,
\end{align}
with $\cal \bar S$ the complement of the subspace $\cal S$. The subspace $\cal S$ and  $\cal \bar S$ contain respectively the initial systems 1 and 2 at the initial time. $\hat U^{TDPHFB}(t_0,t)$ is the  propagator between time $t_0$ and time $t$ of a projected state. The self-consistent propagation of a quasi-particle states would be a theory interesting to  develop. Nevertheless in the present calculation, we choose to propagate independently each states with different gauge angles, with the TDHFB propagator ${\hat U}^{\rm TDHFB}$,
\begin{align}
&\hat U(t_0,t_f)  \hat P_{\cal S}(N_i)  | \Psi(t=0) \rangle \nonumber \\
&\quad = \frac{1}{2 \pi}  \int_0^{2\pi} d \varphi e^{ -i \varphi N_i }    \hat U^{\rm TDHFB}(t_0,t_f)   e^{ i \varphi \hat N_i } | \Psi(t=0) \rangle , \\
&\quad = \frac{1}{2 \pi}  \int_0^{2\pi} d \varphi e^{ -i \varphi N_i }   | \Psi (\varphi,t ) \rangle ,
\end{align}
with $| \Psi (\varphi,t ) \rangle$ the evolved state using the TDHFB equation of motion of the initially state rotated by an angle $\varphi$. 
Using the properties, 
\begin{align}
\hat P_{\cal S}(N_1) \hat P_{\cal \bar S}(N_2) &= \hat P_{\cal S}(N_1) \hat P(N_1+N_2) , \\
e^{i \varphi \hat N} \hat U^{\rm TDHFB}(t_0,t)  &=   \hat U^{\rm TDHFB}(t_0,t) e^{i \varphi \hat N},
\end{align}
we find the expression of the probability,
%
\begin{align}
	P_{\cal S}(N,t)& = \frac{1}{\cal N} \frac{1}{(2\pi)^4} \iiiint_{0}^{2 \pi}     e^{  i (\varphi_1-\varphi_4) N_i -  i \varphi_2 N -i \varphi_3 N_{\rm tot}  }  \nonumber \\
	& \times  \langle \Psi (\varphi_1,t)  |   e^{ i \varphi_2 \hat N_{\cal S}  }  e^{ i \varphi_3 \hat N_{\rm tot}  }       |   \Psi(\varphi_4,t) \rangle d\varphi_{1,2,3,4} ,	\label{eq:four_int}
\end{align}
with the norm,
\begin{align}
	{\cal N }  =   \frac{1}{(2\pi)^3}  &  \iiint_{0}^{2 \pi}    e^{  i (\varphi_1-\varphi_4) N_i   -i \varphi_3 N_{\rm tot}   }    \nonumber  \\ 
	&  \times \langle \Psi(\varphi_1,t)  |      e^{ i \varphi_3 \hat N_{\rm tot}  }       |   \Psi (\varphi_4,t) \rangle d\varphi_{1,3,4} .
\end{align}
The overlap is computed with the Pfaffian method \cite{Ber12},
\begin{align}
	\langle & \Psi(\varphi_1,t)  |   e^{ i \varphi_2 \hat N_{\cal S}  }  e^{ i \varphi_3 \hat N_{\rm tot}  }       |   \Psi (\varphi_4,t) \rangle = 
		(-1)^n \frac{\det C^* \det C'}{\prod_\alpha^n v_\alpha v'_\alpha} {\rm pf} {\cal M} , \\
		{\cal M } &= \left[     
		\begin{array} {cc}
			V^TU & V^T e^{i \varphi_3}(1+\Theta(z) e^{i \varphi_2} ) V'^* \\
			-V'^\dagger  e^{i \varphi_3}(1+\Theta(z) e^{i \varphi_2} ) V & U'^\dagger V'^*
		\end{array}
		    \right].
\end{align}
With C the matrix obtained from the Bloch-Messiah decomposition \cite{RS} and $v_\alpha$ the occupation numbers in the canonical basis. The $C$, $v_\alpha$, $V$ and $U$  correspond  to the bra $ \langle \Psi(\varphi_1,t)  | $   while the $C'$, $v'_\alpha$, $V'$ and $U'$  refer to the ket  $|   \Psi (\varphi_4,t) \rangle$.

\begin{figure}[!ht]
\centering
\includegraphics[width= 0.9 \linewidth]{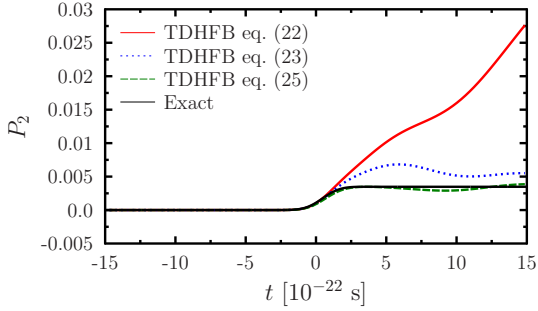}
\caption{  Pair transfer probability determine using several prescription of the TDHFB equation compared to the exact determination of the pair transfer probability.  }
\label{fig:eom}      
\end{figure}

To test this method, we used a slightly different model than previously, we introduce the single particle energy. The four states of each fragments have an energy $e_i$=$i$ MeV with $i$=1,4. Because we use the projection method, we don't take an effective interaction, we take the case of $G=-1$ MeV, $V_0$=-0.03 MeV and $a$=0.3$\times 10^{44} $s$^{-2}$.
Using the standard TDHFB equation,
\begin{align}
  i \hbar \frac{\partial}{\partial t} 
    \left(
    \begin{array}{c}
                U(t) \cr
                V(t)
             \end{array}
             \right)
     = {\cal H}\left( 
    \begin{array}{c}
                U(t) \cr
                V(t)
             \end{array}
             \right), 
\end{align}
with
\begin{align}
  {\cal H}= \left( 
             \begin{array}{cc}
               h& \Delta\cr
               \Delta^{*} & - {h}^{*} 
             \end{array}             
             \right),
             \label{test}
\end{align}
we found a spurious behavior. As figure \ref{fig:eom} shows with red solid line, after the reaction happens, when $V(t)$ become small, the pair transfer probability $P_2 = P_{{\cal S}_1}(6,t)$ is not constant. This is a non-physical behavior that is not present in the exact calculation. We believe that this behavior is due to the approximation that the TDHFB are independently propagate in time. This induce different rotation velocity in the gauge angle plane due to the different chemical potential after the Josephson transfer  take place. To reduce this spurious behavior, our prescription is to remove the chemical  potential in each fragments,
\begin{align}
  {\cal H}= \left( 
             \begin{array}{cc}
               h - \lambda_{\cal S}(t) & \Delta\cr
               \Delta^{*} & - {h}^{*} + \lambda_{\cal S}(t)
             \end{array}             
             \right), 
\end{align}
with $\lambda_{\cal S}$ the chemical potential computed in the subspace $\cal S$ as,
\begin{align}
\lambda_{\cal S} = \frac{1}{\Omega} \sum_{k>0}^{\Omega} {\rm Real} \left(  \frac{\Delta_k(2 n_k - 1 )}{2 \kappa_k} + e_k \right).
\end{align}
This prescription improves the result on fig. \ref{fig:eom}, but a better agreement with the exact solution is found if one remove also the rotation due to the different quasi-particles energy,
\begin{align}
  {\cal H}= \left( 
             \begin{array}{cc}
               h - \lambda_{\cal S}(t) -\epsilon_k(t) & \Delta\cr
               \Delta^{*} & - {h}^{*} + \lambda_{\cal S}(t)-\epsilon_k(t)
             \end{array}             
             \right),
\end{align}
with the quasi-particle energy, 
\begin{align}
\epsilon_{k} = ( e_k - \lambda_{\cal S} ) ( 1 - 2 n_k ) + \Delta_k \kappa^*_k + \Delta^*_k \kappa_k.
\end{align}

Using this TDHFB equation of motion, on fig. \ref{fig:eom}, the transfer probability is almost stable after the two systems are isolated and the average result reproduce well the exact calculation.

\section{Conclusion}

In conclusion, we used a simple model of reaction between two superfluid systems. We interpret the fluctuations of the TDHFB observables with respect to the initial relative gauge angle as statistical fluctuations. The average value and the standard deviation of the energy and the number of transfered particles from TDHFB are closed to the exact results. We then developed a projection method to determine the transfer probabilities. A spurious behavior is found after the separation of the two systems, a prescription to modify the TDHFB equation in order to cure the problem is proposed and reproduce correctly the exact results. This prescription will be used in the case of a realistic calculation in a future contribution.

\end{document}